\documentclass[conference]{IEEEtran}
\IEEEoverridecommandlockouts
\usepackage{cite}
\usepackage{amsmath,amssymb,amsfonts}
\usepackage{algorithmic}
\usepackage{graphicx}
\usepackage{textcomp}
\usepackage{hyperref}
\usepackage{xcolor}
\usepackage{booktabs}
\usepackage{multirow}
\hypersetup{
    colorlinks=true,
    linkcolor=blue,
    citecolor=blue,
    urlcolor=blue
}
\usepackage[linesnumbered,ruled,vlined]{algorithm2e}
\def\BibTeX{{\rm B\kern-.05em{\sc i\kern-.025em b}\kern-.08em
    T\kern-.1667em\lower.7ex\hbox{E}\kern-.125emX}}

\usepackage{subcaption}
\usepackage{fancyhdr}

\begin{document}
\title{SEAFL: Enhancing Efficiency in Semi-Asynchronous Federated Learning through Adaptive Aggregation and Selective Training
\\
} 
\author{
    \IEEEauthorblockN{Md Sirajul Islam\textsuperscript{1}, Sanjeev Panta\textsuperscript{1}, Fei Xu\textsuperscript{2}, Xu Yuan\textsuperscript{3}, Li Chen\textsuperscript{1}, and Nian-Feng Tzeng\textsuperscript{1}}
    \IEEEauthorblockA{\textsuperscript{1}School of Computing and Informatics, University of Louisiana at Lafayette, USA\\
    \textsuperscript{2}School of Computer Science and Technology, East China Normal University, China\\
    \textsuperscript{3}Department of Computer and Information Sciences, University of Delaware, USA }
    \thanks{The research is supported in part by the NSF under grants OIA-2327452 and OIA-2019511, in part by the Louisiana BoR under LEQSF(2024-27)-RD-B-03, and in part by the NSFC under 62372184 and by the Sci. and Tech. Commission of Shanghai Municipality under 22DZ2229004.}
    \thanks{Corresponding author: Li Chen. Email: li.chen@louisiana.edu}
}

\maketitle

\begin{abstract}
Federated Learning (FL) is a promising distributed machine learning framework that allows collaborative learning of a global model across decentralized devices without uploading their local data. However, in real-world FL scenarios, the conventional synchronous FL mechanism suffers from inefficient training caused by slow-speed devices, commonly known as stragglers, especially in heterogeneous communication environments. Though asynchronous FL effectively tackles the efficiency challenge, it induces substantial system overheads and model degradation. Striking for a balance, semi-asynchronous FL has gained increasing attention, while still suffering from the open challenge of stale models, where newly arrived updates are calculated based on outdated weights that easily hurt the convergence of the global model. In this paper, we present {\em SEAFL}, a novel FL framework designed to mitigate both the straggler and the stale model challenges in semi-asynchronous FL. {\em SEAFL} dynamically assigns weights to uploaded models during aggregation based on their staleness and importance to the current global model. We theoretically analyze the convergence rate of {\em SEAFL} and further enhance the training efficiency with an extended variant that allows partial training on slower devices, enabling them to contribute to global aggregation while reducing excessive waiting times. We evaluate the effectiveness of {\em SEAFL} through extensive experiments on three benchmark datasets. The experimental results demonstrate that {\em SEAFL} outperforms its closest counterpart by up to $\sim$22\% in terms of the wall-clock training time required to achieve target accuracy. 

\end{abstract}


\section{Introduction}
\label{sec:introduction}
In recent years, the proliferation of edge devices has resulted in a significant surge in distributed data generation that can be leveraged for machine learning and smart applications. However, with the introduction of stringent laws and regulations such as the GDPR \cite{gdpr} in 2018, traditional methods based on data aggregation into a centralized data center raise serious privacy concerns and become increasingly unfeasible. As a promising alternative, Federated Learning (FL) \cite{mcmahan2017communication} has emerged to enable collaborative model training without the need of transferring raw data. FL leverages distributed user data while preserving privacy by exchanging the gradients or model updates of participating devices. Due to its superior privacy implications, FL has been applied in diverse areas such as natural language processing \cite{liu2021federated}, computer vision \cite{liu2020fedvision}, healthcare \cite{nguyen2022federated}, and human activity recognition \cite{ouyang2021clusterfl}.

Traditional FL \cite{mcmahan2017communication} training typically relies on a parameter server to orchestrate the training process across devices using a synchronous mechanism. This synchronous training approach involves multiple rounds, each comprising the following steps. Initially, the server chooses a subset of devices and broadcasts the global model to them. Then, local training is performed on each selected device using its own data. Subsequently, each device sends the model updates back to the server. Finally, the server aggregates the received updates to produce a new global model once all chosen devices finish the aforementioned steps. Despite its efficiency and ease of implementation, the synchronous mechanism is susceptible to stragglers (slow devices), which can significantly prolong the training process \cite{jiang2022pisces}, particularly when dealing with heterogeneous devices \cite{lai2021oort, yang2021characterizing}. This could severely impact training efficiency as powerful devices may remain inactive while the server waits for stragglers \cite{wu2020safa}, posing critical challenges that greatly hinder the scalability of synchronous FL methods in large-scale cross-device scenarios.

To tackle these limitations, recent studies have introduced asynchronous FL (AFL) \cite{xie2019asynchronous, nguyen2022federated,wu2020safa,xu2023asynchronous}, allowing the server to aggregate uploaded models without waiting for stragglers, which may instead contribute to future aggregation rounds. In fully AFL such as FedAsync \cite{xie2019asynchronous}, the server initiates the global model aggregation immediately upon receiving a single model update. Although this approach alleviates the straggler issue, it introduces stale model updates, leading to slower convergence and accuracy degradation \cite{zhou2022efficient}. Additionally, it incurs significant computational overhead due to excessive server aggregation. 

\setlength{\abovecaptionskip}{.5pt}
\begin{figure*}
\centering
{\includegraphics[height=4cm, width=0.94\textwidth]{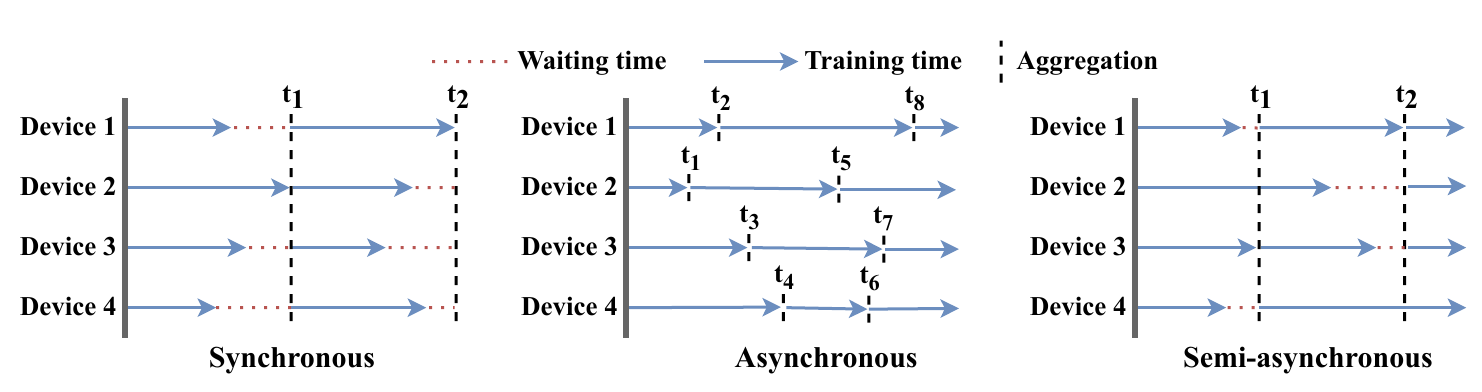}}

\caption{The working process of synchronous, asynchronous, and semi-asynchronous FL algorithms.}
\label{working}
\end{figure*}

As a compromise, semi-asynchronous FL methods \cite{nguyen2022federated, wu2020safa, zhou2024asynchronous, liu2023fedasmu} buffer a specified number of local updates for aggregation in each round, as illustrated in Fig.~\ref{working}. 
Once receiving a sufficient number ({\em i.e.}, 3 for the example in Fig.~\ref{working}) of updates, the server updates the global model without waiting for slower devices ({\em i.e.}, Device 2 in the first round). Those devices failing to participate in the aggregation can continue their training to completion and potentially contribute to future aggregation.
Some approaches \cite{wu2020safa, liu2023fedasmu} discard local updates from slower devices based on a staleness threshold, resulting in wasted training efforts. 
Excluding them would also impede the convergence of the global model and delay the training process. To let slower devices contribute to the global aggregation while accounting for the staleness of their updates, existing semi-asynchronous FL methods leverage static polynomial formulas \cite{wang2022asyncfeded, nguyen2022federated} or simple attention mechanisms \cite{chen2020asynchronous}. However, they are limited in their abilities to determine and dynamically adjust the significance of received updates during aggregation, resulting in suboptimal training efficiency and model accuracy. 



To fill this gap, we propose a novel staleness-aware semi-asynchronous FL framework ({\em SEAFL}), to effectively and efficiently learn from devices with heterogeneous system characteristics. Based on empirical insights, {\em SEAFL} strikes for an optimal balance between involving more devices to contribute to global aggregations and reducing aggregation overheads. Moreover, having identified that the contribution of each device's local updates on the global model varies with their staleness across rounds, {\em SEAFL} dynamically assigns weights to local updates during aggregation to ensure efficient collaborative learning in heterogeneous environments. The essence of {\em SEAFL} lies in an adaptive weight aggregation mechanism to address the stale model problem by considering both the staleness of the received model updates and their similarity to the current global model. In contrast to prior work, our method emphasizes the importance of local updates according to the current global model which effectively accelerates model convergence. Additionally, to enhance training efficiency, we introduce a variant,  {\em SEAFL$^2$}, which further reduces the training time by enabling partial training on straggler devices. To demonstrate the efficacy of {\em SEAFL}, we conduct extensive experiments on three benchmark datasets, comparing our approach with existing state-of-the-art (SOTA) FL methods. Results demonstrate that {\em SEAFL} significantly outperforms the SOTA FL approaches, especially its closest counterpart, {\em FedBuff}, by reducing the wall-clock training time required to achieve target accuracy, for up to $\sim$22\%.

Our key contributions are summarized as follows:
\begin{itemize}
\item We investigate the impacts of local update staleness, buffer size, and importance of local updates in asynchronous FL training.
\item We introduce {\em SEAFL}, a novel staleness-aware semi-asynchronous FL framework that adaptively assigns weights to local updates while considering their staleness and importance to the current global model.
\item We empirically show the effectiveness of {\em SEAFL}, which outperforms other SOTA FL methods in terms of reducing the wall-clock training time required to achieve target accuracy.
\item We present a theoretical analysis of the convergence behavior of the proposed {\em SEAFL} algorithm. 
\item In addition, to improve training efficiency, we propose a variant of {\em SEAFL}, {\em SEAFL$^2$}, that enables partial training on slower devices, allowing them to contribute to the global aggregation process and to reduce the excessive waiting time.
\end{itemize}

The rest of this paper is structured as follows. We review related work in Section~\ref{sec:relatedwork}. Section~\ref{sec:pre} provides our preliminary insights. Section~\ref{sec:design} outlines the problem formulation and the design of our proposed {\em SEAFL} framework. A theoretical convergence analysis is presented in Section~\ref{sec:theory}. The experimental settings and results are given in Section~\ref{sec:results}. Finally, Section~\ref{sec:end} concludes the paper.

\section{Related Work}
\label{sec:relatedwork}

\subsection{Synchronous Federated Learning}
A plethora of FL approaches \cite{mcmahan2017communication} have been proposed to jointly train a global model by leveraging distributed user data. Many of them \cite{mcmahan2017communication, li20201federated} rely on a synchronous mechanism for aggregating models on the server. However, this approach requires the server to wait for all selected devices to transmit their model updates before performing aggregation, which has proven inefficient due to the presence of stragglers. As the number of devices increases and system heterogeneity grows, the probability of encountering straggler effects also rises. This issue significantly impedes the scalability of synchronous FL. Existing work tackles system heterogeneity and statistical heterogeneity separately. Several approaches, including regularization \cite{li20201federated}, personalization \cite{li2021ditto, zhang2023fedala}, clustering \cite{ghosh2020efficient, islam2024fedclust}, and device selection \cite{lai2021oort, javaherian2024fedfair}, have been proposed in the literature to tackle statistical heterogeneity. However, these approaches lack the capability to dynamically adjust the significance of diverse models and instead focus solely on the synchronous mechanism.

Three different strategies are introduced in the literature to tackle system heterogeneity within the synchronous mechanism. Firstly, some methods focus on scheduling appropriate devices for local training while considering their computational and communication capabilities to achieve load balance and mitigate inefficiencies caused by stragglers \cite{lai2021oort, li2022pyramidfl}. However, this type of approaches may reduce the participation frequency of less powerful devices, leading to decreased accuracy. Secondly, techniques such as pruning \cite{zhang2022fedduap} or dropout \cite{horvath2021fjord} are leveraged during training, resulting in lossy compression and reduced accuracy. Thirdly, the clustering approach \cite{li2022fedhisyn} groups devices with similar capacities into clusters and utilizes a hierarchical architecture \cite{abad2020hierarchical} for model aggregation. Although these approaches aim to optimize the synchronous mechanism, they often suffer from low efficiency and may lead to significant accuracy degradation due to statistical heterogeneity. 
\subsection{Asynchronous and Semi-asynchronous FL}
To address the system heterogeneity, AFL \cite{nguyen2022federated, xie2019asynchronous} facilitates global model aggregation without the need to wait for all devices. In AFL, aggregation can be performed immediately upon receiving an update from any device \cite{xie2019asynchronous,chen2020asynchronous} or when multiple updates are buffered \cite{nguyen2022federated,liu2023fedasmu,zhang2023timelyfl}. In {\em FedAsync} \cite{xie2019asynchronous}, the server employs a mixing hyperparameter $\alpha$ to determine the weight allocated to the newly arrived model update based on that of the fastest device during the aggregation. In fully AFL \cite{xie2019asynchronous, chen2020asynchronous}, the aggregation process is no longer delayed by slower devices. Upon finishing their local training, their model updates may be based on an earlier version of the global model compared to those of faster devices. However, outdated uploaded models from stale devices may revert the global model to a previous state, significantly reducing accuracy \cite{chen2020asynchronous}. Furthermore, it incurs excessive computation overhead due to frequent aggregation on the server.

Hence, the semi-asynchronous FL was introduced as a trade-off between synchronous and asynchronous FL. It alleviates the excessive computation overhead and privacy concerns by buffering a certain number of local updates instead of aggregating them immediately. Wu \textit{et al.} proposed {\em SAFA} \cite{wu2020safa}, which categorizes devices according to their training status to enhance convergence performance. It discards stale model updates based on a hyperparameter called lag tolerance. {\em FedSA} \cite{ma2021fedsa} introduced a two-phase FL training process, employing a large number of epochs during the initial training phase, and then switching to a reduced number of local epochs in the convergence phase. It adjusted the number of local training epochs in each round according to the device's staleness. {\em Fedbuff} \cite{nguyen2022federated} enables secure aggregation by keeping a predefined number of local updates in a secure buffer before aggregation. Liu \textit{et al.} proposed {\em FedASMU}, \cite{liu2023fedasmu} a reinforcement learning approach to dynamically choose a time slot for triggering server-side aggregation. However, it incurs additional computation overhead on both the device and server side. 

Recent work, {\em EAFL} \cite{zhou2024asynchronous}, introduced gradient similarity-based clustering and a two-stage aggregation strategy to address data and system heterogeneity issues in asynchronous FL. Nevertheless, it relies on a predefined number of clusters, which is challenging to determine without knowing the actual data distributions across devices, thus limiting flexibility and adaptability.  Most of prior pursuits \cite{nguyen2022federated, xie2019asynchronous} did not impose any staleness limitations on device updates, resulting in stale model updates that hinder the convergence of the final model. Furthermore, it does not perform well in cases of low data heterogeneity while incurring additional computation overhead. Unlike existing approaches, we introduce a semi-asynchronous FL framework, i.e., {\em SEAFL}, to tackle system heterogeneity. {\em SEAFL} dynamically adjusts weights to the received model updates according to their staleness and importance during global aggregation to minimize loss and improve accuracy. Moreover, our approach facilitates partial training on slower devices, enabling them to contribute to the global aggregation. 

\section{Preliminary Insights}
\label{sec:pre}
In this section, we conduct preliminary experiments to analyze the impact of buffer size, model staleness, and the significance of local model updates to the global aggregation on semi-asynchronous FL training.

\begin{figure*}
\centering
\captionsetup[subfigure]{skip=-3pt}
\subfloat[Varying the buffer size (K)]{%
   \includegraphics[height=5.2cm, width=0.33\textwidth]{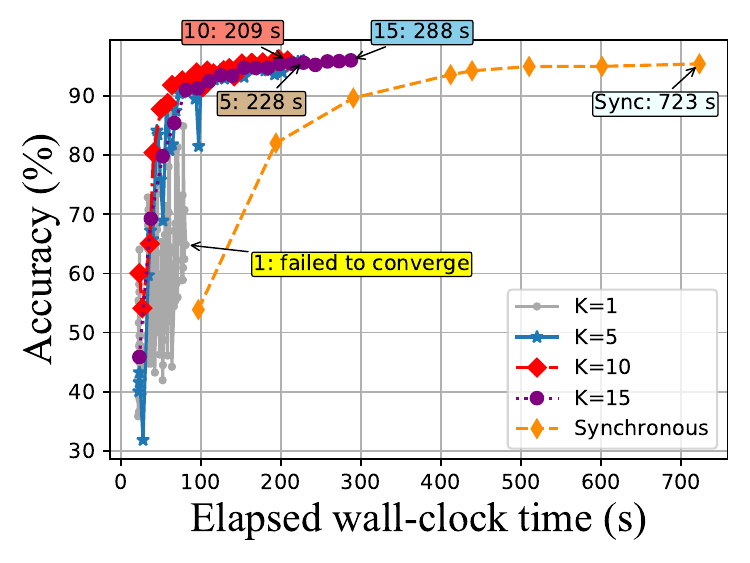}%
   \label{fig:subfig:buffer}}
\subfloat[Varying the limit of staleness ($\beta$)]{%
   \includegraphics[height=5.2cm, width=0.33\textwidth]{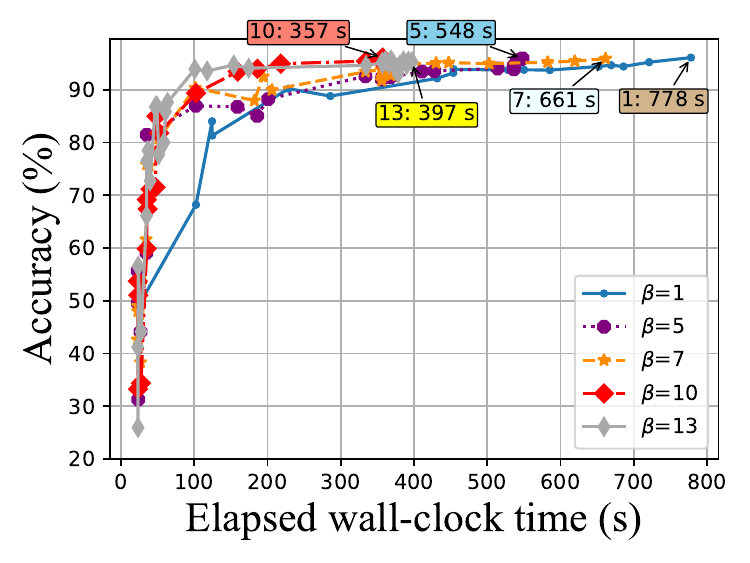}%
   \label{fig:subfig:staleness}}
\subfloat[Effects of local updates importance]{%
   \includegraphics[height=5cm, width=0.33\textwidth]{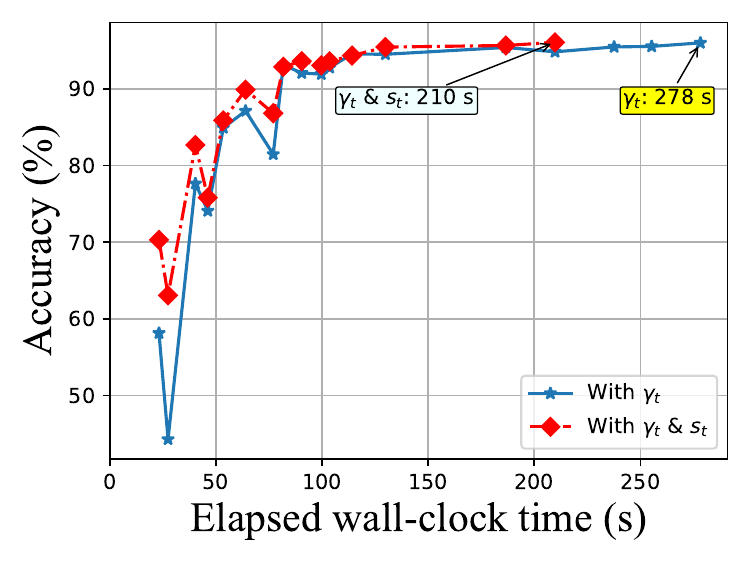}%
   \label{fig:subfig:local}}
\caption{Illustration of the impacts of buffer size, staleness limit, and importance of local updates on asynchronous FL, where ${\gamma}_t$ indicates staleness factors, and $s_t$ denotes the importance of updates.}
\label{figure2}
\end{figure*}

Our experimental setup involves 100 devices utilizing the MNIST dataset to train a LeNet-5 model. We simulate a non-IID distribution using the Dirichlet distribution \cite{li2022federated} with a concentration parameter 0.3. Each device trains the model using 600 training samples. AFL is most suitable for scenarios where a few devices exhibit significantly slower training speeds, leading to a heavy-tailed distribution of local training speeds. To simulate this scenario in our testbed, we randomly generate idle period durations for each device after completing an epoch. These durations are sampled from a Zipf distribution \cite{li2023plato} with parameter \textit{s} = 1.7 and a maximum length of 60 seconds. In the synchronous mode, the server chooses 20 devices for training in each round. We vary the values of buffer size (\textit{K}) and staleness limit ($\beta$) in a semi-asynchronous FL setting, and plot the results in Fig.~\ref{figure2}. We have the following observations.

\setlength{\abovecaptionskip}{1pt}

\textbf{Buffer size.} In AFL, the buffer size refers to the number of updates the server will wait before triggering the aggregation process. When the buffer size is set to 1, the server operates in fully asynchronous mode and immediately performs aggregation upon receiving any update. For instance, {\em FedAsync} \cite{xie2019asynchronous} and {\em ASO-Fed} \cite{chen2020asynchronous} are designed to work in this fully asynchronous mode. On the contrary, when the buffer size is equal to the number of devices selected for each round, the training reverts to the synchronous mode, similar to traditional {\em FedAvg} \cite{mcmahan2017communication}. In this scenario, the server waits for all the chosen devices to upload their updates before starting the aggregation process.

We measure the elapsed wall-clock time as a performance metric while varying the number of updates needed before the server starts aggregation. Fig.~\ref{fig:subfig:buffer} clearly illustrates that the fully asynchronous approach, which immediately commences aggregation upon the arrival of a single update, failed to achieve convergence. This is because each device uses only 600 samples, and the server frequently aggregates model updates from faster devices. Consequently, when updates from considerably slower devices finally arrive, they are based on significantly outdated models. Moreover, the presence of a non-IID data distribution further exacerbates this issue.

On the other hand, synchronous FL did achieve convergence; however, it required a significantly longer wall-clock time. This indicates the well-known straggler issue, where the server is forced to wait for slower devices in each round. In our experiments, aggregating a minimum of 10 device updates yields the optimal outcome, taking only 209 seconds to reach a target accuracy of 96\%. 


\textbf{Model staleness.}
The staleness of devices refers to the number of rounds that have passed since the device last received the global model from the server. We vary the staleness limit to observe its impact on the wall-clock time required to reach convergence. Intuitively, it may not be ideal to impose excessive restrictions when waiting for devices that are only slightly behind. Conversely, we must also be cautious not to incorporate devices that are excessively stale, as their models may be significantly out of sync with the majority. As depicted in Fig.~\ref{fig:subfig:staleness}, the results of our preliminary experiments conducted on the MNIST dataset with \textit{K} = 10 appear to confirm our intuition, indicating that the staleness limit of 10 provides the best performance. The notable difference in performance reveals that achieving a target accuracy of 96\% required 778 seconds with a staleness limit of 1, whereas it only took 357 seconds with a staleness limit of 10.
The choice of staleness limit significantly impacts the wall-clock time required to reach a target accuracy in asynchronous FL. 

\textbf{Importance of uploaded models.}
In AFL, multiple devices are training simultaneously using different versions of the global model. Their updates, when used for aggregation, may not equally contribute to or even be beneficial for the global model convergence. We measure the importance of each received update relative to the current global model, to be outlined in Section IV. Intuitively, if the aggregation weight of each update is set to be proportional to the contribution of the device, the performance can be further improved. Fig.~\ref{fig:subfig:local} illustrates that incorporating the significance of local updates reduces the wall-clock time to achieve the target accuracy to 210 seconds, compared to 278 seconds without this consideration.

\textbf{\textit{Insight:}} The performance of semi-asynchronous FL is greatly impacted by the buffer size and the stateless limit. In addition, not all local updates contribute equally for global optimization, calling for a weighting scheme that adaptively assigns weights to received updates based on their significance degrees for global aggregation.

Inspired by the aforementioned observations and insights, we have designed a staleness-aware semi-asynchronous FL framework with an adaptive weight aggregation mechanism called {\em SEAFL}. It dynamically assigns weights to received updates based on their staleness and importance degrees. The specifics of {\em SEAFL} will be described in the next section.

\begin{table}[t]
\caption{Notations and Descriptions}
\centering
\begin{tabular}{ll}
\hline
\textbf{Symbol} & \textbf{Description} \\
\hline
\textit{\textit{D}} & the complete dataset\\
\textit{$\mathcal{D}_k$} & the local dataset of device \textit{k}\\

\textit{N} & the number of devices \\
\textit{K} & the buffer size of device updates \\
\textit{E} & the number of local training epoch \\
\textit{t} & the current communication round\\
\textit{$S_k$} & the staleness of device \textit{k}'s update\\
\textit{$\alpha$} & staleness weight\\
\textit{$\gamma_t^k$} & the staleness factor for device \textit{k}'s update at round $t$\\
\textit{$s_t^k$} & the importance of device \textit{k}'s update at round $t$\\
\textit{$\mu$} & similarity weight\\
\textit{$\beta$} & staleness limit\\
\textit{$\Theta$} & the cosine similarity between two vectors\\
\textit{$p_k$} & the weight assign to device \textit{k} updates during aggregation\\
\hline
\end{tabular}
\label{table1}
\end{table}

\section{System Design}
\label{sec:design}
\subsection{Problem Formulation}
In this section, we present the formulation of the FL training problem in a simplified setting. For clarity, we provide a list of key notations frequently used throughout this paper in Table~\ref{table1}. Consider a group of \textit{N} devices collaborating in a federated learning process to train a shared model and determine an optimal set of parameters that minimize the global loss function:
\begin{equation}
\min_{w} F(w) \overset{\scriptscriptstyle\Delta}{=} \sum_{k=1}^{N} p_kF_k(w)
\end{equation}

Here, device \textit{k} has a local dataset $\mathcal{D}_k$, $D =\sum_{k=1}^{N} \mathcal{D}_k$, and $p_k =\frac{|\mathcal{D}_k|}{|D|}$.
Subsequently, the local objective function of device \textit{k} is defined as the empirical loss computed over its local dataset, $\mathcal{D}_k$: 

\begin{equation}
F_k(w)=\frac{1}{|\mathcal{D}_k|}\sum_{j_k=1}^{\mathcal{D}_k}f_{j_k} (w; x_{j_k},y_{j_k})
\end{equation}
where $|\mathcal{D}_k|$ represents the number of local samples on each device. Each device trains the model independently on its local dataset and transmits the model updates back to the server. The most widely used synchronous FL algorithm {\em FedAvg} \cite{mcmahan2017communication} aggregates received local model updates after each round to produce the new global model as: 
\begin{equation}
w_{t+1}^{{g}} \leftarrow \sum_{k=1}^{M} p_kF_k(w)
\end{equation}

Here, \textit{M} is the number of devices selected for training in each round. It naively assigns weights ($p_k$) to device updates while aggregating based on the percentage of each device samples among the total number of samples in each round. However, this straightforward weight allocation scheme fails to account for the staleness and significance of model updates in asynchronous FL training scenarios. Consequently, it leads to a degradation in the accuracy of the global model, especially when dealing with stragglers and non-IID data distributions.

\subsection{Adaptive Weight Aggregation}
In this section, we present the proposed {\em SEAFL} with an adaptive weighted aggregation mechanism that dynamically allocates weight to each received update based on their staleness, and the importance of each update compared to the current global model. The primary design objective of {\em SEAFL} is to optimize the wall-clock training time required for an FL task to achieve a target accuracy, rather than focusing on the total number of communication rounds. We have identified key influential factors affecting AFL training through preliminary experiments and subsequently use these factors to assign weights to received updates adaptively during aggregation. 

\SetKwInput{KwInput}{Input}
\SetKwInput{KwOutputOne}{Output} 
\SetKwInput{KwOutputTwo}{Server Initializes}
\SetKwInput{KwOutputThree}{Server Executes}  
\SetKwInput{KwOutputFour}{ClientUpdate} 
\SetAlgoNoEnd
\SetAlgoNoLine
\begin{algorithm}[t]

\DontPrintSemicolon
  \KwInput{\textit{N}: Number of available clients, \textit{K}: buffer size, $\beta$: staleness limit, \textit{E}: local training epochs,
  $\eta$: local learning rate, \textit{B}: local mini-batch size, $\alpha$: staleness factor , $\mu$: similarity factor.}
  \KwOutputOne{$w_T$: The global model at Round \textit{T};\\
                }
  \KwOutputTwo{Initialize \textit{t} = 0, $w_0^g$;\\
                }
  \KwOutputThree{\\
  \For{t = 0, 1, 2,..., $T-1$}{
        Server chooses a subset $\mathcal{S}_t$ of \textit{N} devices at random;\\
        Broadcast $w_t^g$ to all selected clients;\\
        flag = 0;\\
        \While{${flag}\leq K$}{
            Server receives local model updates from clients $w_t^k$;\;
            Server stores received updates into the buffer;\;
            $flag$ +=1;\;
            }
        \textbf{end}\;
        {
        Server evaluates $\gamma_t^k$ by Eq. (4);\; 
        Server calculates $s_t^k$ by Eq. (5);\; 
        Server determines $p_t^k$ by Eq. (6);\; 
        Server aggregates parameters in \textit{K}:\;
        \quad $w_{t+1}^g \leftarrow \sum_{k=1}^{K} p_t^kw_t^k$ ;\;
        Server updates the global model:\;
        \quad $w_{t+1}^{g} \leftarrow (1-\vartheta)w_{t}^{g} + \vartheta w_{t}^{new}$;\;
        Server sends $w_{t+1}^g$ to the \textit{K} newly updated clients;
        }
        }
        \textbf{end}\;
        }
        
  \KwOutputFour{\\
  \qquad Client \textit{k} receives global model parameter $w_t^g$;\;
  \qquad $w_t^k\leftarrow w_t^g$;\;
  \qquad \For{each client $k \in \mathcal{S}_t$ in parallel}{
  \qquad \For{each local epoch $l = 1, 2, \ldots, E$}{
  \qquad \For{each batch $b$ in $B_k$}{
        \qquad $w_{t+1}^k = w_{t}^k - \eta \Delta f(w_t^k; b)$;\;
    }
    \qquad \textbf{end}\;
    }
    \qquad \textbf{end}\;
    }
  \qquad \textbf{end}\;
  \qquad Upload $w_{t+1}^k$ to the server;\;
  
  }
\caption{\em {SEAFL}} 
\label{algo1}
\end{algorithm}

\textbf{Staleness factor.} In AFL training, slower devices that obtained the global model from the server several rounds earlier are prone to have outdated updates. As a result, their model updates may not significantly contribute to the aggregation process in terms of quality, resulting in slower convergence of the global model. Therefore, the weight allocated to these outdated updates should be reduced during aggregation. As the staleness of an update increases, its aggregation weight should also be correspondingly diminished. Since {\em SEAFL} synchronously waits for devices that exceed the staleness threshold, their staleness will always remain below that threshold. More specifically, let $t$ denote the ongoing round at the server, and $t_k$ represent the round in which device \textit{k} last obtained its model from the server. The staleness of device \textit{k}'s update is computed as $t-t_k$. We measure the staleness of each update using the following staleness function, which will be used for adjusting the aggregation weights:
\begin{equation}
\gamma_t^k=\alpha\cdot\frac{\beta}{(t-t_k) + \beta}
\end{equation}

Here, $t-t_k = S_k$ represents the staleness of device \textit{k}'s update, $\beta$ is the staleness limit which follows $S_k \leq \beta$, and  $\alpha$ serves as a hyperparameter controlling the significance of the staleness factor in the aggregation process.

\setlength{\abovecaptionskip}{.5pt}
\begin{figure*}
\centering
{\includegraphics[height=6.4cm, width=0.9\textwidth]{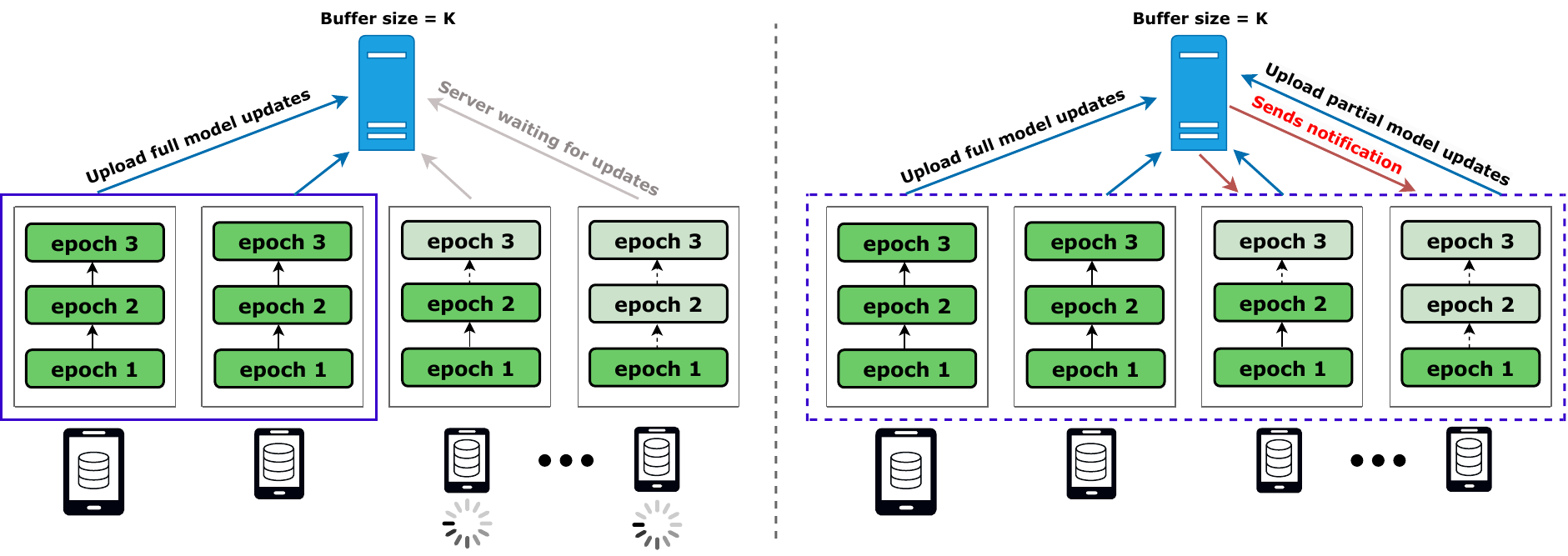}}

\caption{\textbf{Left}: The traditional AsyncFL architecture where the server initiates aggregation upon receiving the required number of local updates. \textbf{Right}: The proposed {\em SEAFL$^2$} allows partial training on slower devices, enabling them to contribute to global aggregation. The server will notify slower devices to send their updates immediately after exceeding the staleness limit.}
\label{figure3}
\end{figure*}

\textbf{Importance of updates.}
From our preliminary experiments, it is evident that considering staleness alone does not yield optimal results. To improve the {\em SEAFL} performance further, we introduce the concept of incorporating the importance of local updates relative to the current global model. Specifically, we prioritize local updates that demonstrate higher similarity to the current global model and consequently assign them a higher weight. Mathematically, two methods can be utilized to assess the similarity between two vectors quantitatively. The first method involves computing the dot product, which considers both the magnitude and the angle between the vectors. In contrast, cosine similarity offers an alternative by focusing exclusively on the angle between the vectors. In {\em SEAFL}, we utilize cosine similarity, represented as $s_t^k$, to measure the similarity between two vectors quantitatively. A lower value of $s_t^k$ indicates less similarity between the two vectors. We normalize $s_t^k$ values to [0, 1] by computing $(\Theta+1)/2$ instead. The significance of the update received from device \textit{k} at global round \textit{t} is therefore defined as: 
\begin{equation}
s_t^k=\mu\cdot\frac{\Theta (\Delta_t^k , w_t^g) + 1}{2}
\end{equation}

Similar to the staleness factor, we introduce another hyperparameter $\mu$, serving as another tuning knob to control the importance of each update during aggregation. After computing both influential factors, and considering that each device \textit{k} executes \textit{E} training epochs on its local dataset $\mathcal{D}_k$, the aggregation weight for each device can be determined as follows:
\begin{equation}
p_t^k = \frac{|\mathcal{D}_k|}{|D|} (\gamma_t^k + s_t^k)
\end{equation}
in which \textit{D} represents the collection of all data samples utilized by the participating devices \textit{K} in the current round. The server normalizes all $p_t^k$ so that their sum equals 1, and then aggregates the \textit{K} parameters from the buffer as follows:
\begin{equation}
w_{t}^{new} \leftarrow \sum_{k=1}^{K} p_t^kw_t^k
\end{equation}

After acquiring $w_{t}^{new}$, the server employs a weighted averaging strategy to update the global model:
\begin{equation}
w_{t+1}^{g} \leftarrow (1-\vartheta)w_{t}^{g}+\vartheta w_{t}^{new}
\end{equation}
where the hyper-parameter $\vartheta\in(0,1)$. The server then transmits the updated global model $w_{t+1}^{g}$ to the newly updated device for the upcoming round of local training. The pseudocode of {\em SEAFL} is shown in Algorithm~\ref{algo1}.

\subsection{Partial Training}
In the design of {\em SEAFL}, we acknowledge that the adaptive weighted aggregation mechanism by itself may not provide optimal outcomes in terms of wall-clock training time, especially when compared to the total number of rounds. This issue arises when a few significantly slower devices exceed the staleness threshold, potentially becoming stragglers and resulting in an extended training time needed to reach the target accuracy. Unlike existing works \cite{nguyen2022federated,liu2023fedasmu} addressing the AFL challenges, which require all devices to complete an equal number of local epochs for each update regardless of device heterogeneity, the proposed {\em SEAFL$^2$} enables partial training on slower devices. Moreover, some existing approaches \cite{wu2020safa, liu2023fedasmu} discard local updates that exceed a predefined staleness limit, resulting in wasted training efforts and slower devices being unable to contribute to the global aggregation which may impede the model's convergence.

To alleviate the negative impacts of these stragglers, the server in {\em SEAFL$^2$} notifies all devices that exceed the staleness limit. Upon receiving such a notification, devices refrain from advancing to the next epoch of local training. Instead, they immediately transmit their local model updates upon completing the ongoing training epoch. This proves beneficial for the server when dealing with slower devices, as it eliminates the need to wait for the completion of all local epochs on those devices. Rather, the server only needs to wait for the ongoing epoch to finish.

Fig.~\ref{figure3} illustrates how the server notifies stale devices to send their updates immediately. The buffer size and staleness limit discussed in Section III remain applicable in {\em SEAFL$^2$}. The server continuously waits to receive a requisite number of device updates. However, with the partial training strategy, it also monitors whether any devices exceed the staleness limit. If such scenarios arise, the server will send notifications to these devices. These notifications introduce an additional round trip between the server and devices with stale updates. After receiving the notification, these devices will transmit their model updates immediately upon finishing their current training epoch.

\SetKwInput{KwInput}{Input}
\SetKwInput{KwOutputOne}{Output} 
\SetKwInput{KwOutputTwo}{Server Initializes}
\SetKwInput{KwOutputThree}{Server Executes}  
\SetKwInput{KwOutputFour}{ClientUpdate} 
\SetAlgoNoEnd
\SetAlgoNoLine
\begin{algorithm}[t]

\DontPrintSemicolon
  \KwInput{\textit{N}: Number of available clients, \textit{K}: buffer size, $\beta$: staleness limit, \textit{E}: local training epochs,
  $\eta$: local learning rate, \textit{B}: local mini-batch size, $\alpha$: staleness factor , $\mu$: similarity factor.}
  \KwOutputOne{$w_T$: The global model at Round \textit{T};\\
                }
  \KwOutputTwo{Initialize \textit{t} = 0, $w_0^g$;\\
                }
  \KwOutputThree{\\
  \For{t = 0, 1, 2,..., $T-1$}{
        Server chooses a subset $\mathcal{S}_t$ of \textit{N} devices at random;\\
        Broadcast $w_t^g$ to all selected clients;\\
        flag = 0;\\
        \While{${flag}\leq K$}{
            Server receives local model updates from clients $w_t^k$;\;
            Server stores received updates into the buffer;\;
            $flag$ +=1;\;
            }
        \textbf{end}\;
        \For{each client $k \in \mathcal{S}_t$}{
            \If{client $k$'s update exceed $\beta$}{
                Send a notification to client $k$;\;
             }
        }
        \textbf{end}\;
        {
        Server evaluates $\gamma_t^k$  by Eq. (4);\;
        Server calculates $s_t^k$  by Eq. (5);\; 
        Server determines $p_t^k$  by Eq. (6);\; 
        Server aggregates parameters in \textit{K}:\;
        \quad $w_{t+1}^g \leftarrow \sum_{k=1}^{K} p_t^kw_t^k$ ;\;
        Server updates the global model:\;
        \quad $w_{t+1}^{g} \leftarrow (1-\vartheta)w_{t}^{g} + \vartheta w_{t}^{new}$;\;
        Server sends $w_{t+1}^g$ to the \textit{K} newly updated clients;
        }
        }
        \textbf{end}\;
        }

  \KwOutputFour{\\
  \qquad Client \textit{k} receives global model parameter $w_t^g$;\;
  \qquad $w_t^k\leftarrow w_t^g$;\;
  \qquad \For{each client $k \in \mathcal{S}_t$ in parallel}{
  \qquad \For{each local epoch $l = 1, 2, \ldots, E$}{
  \qquad \For{each batch $b$ in $B_k$}{
        \qquad $w_{t+1}^k = w_{t}^k - \eta \Delta f(w_t^k; b)$;\;
    }
    \qquad \textbf{end}\;
    \qquad \If{$k$ receives a notification}{ 
      \qquad Finish the current epoch;\;
      \qquad Send $w_{t+1}^k$ to the server immediately;\;
    }
    \qquad \Else{
      \qquad Continue training remaining epochs;\;
        }
    }
    \qquad \textbf{end}\;
    }
  \qquad \textbf{end}\;
  \qquad Upload $w_{t+1}^k$ to the server;\;
}
\caption{\em {SEAFL$^2$}}
\end{algorithm}

\section{Theoretical Analysis}
\label{sec:theory}
We consider the following theoretical context to analyze {\em SEAFL}'s convergence behavior. In each round $t \in T$, the server chooses \textit{M} devices from a pool of \textit{N} devices. Each device \textit{k} executes \textit{E} epochs of training on its local dataset $\mathcal{D}_k$, utilizing the model $w_{t_k}^{k}$ received from the server in round $t_k$. During each local training epoch $i\in [0, E]$, the local model $w_{t_k, i+1}^{k}$ is updated using an SGD optimizer with a learning rate $\eta_l^i$ and a batch size \textit{B}. This process can be expressed as $w_{t_k, i+1}^{k} = w_{t_k}^k - \eta_l^i g(w_{t_k, i}^k)$, where the gradient $g(w_{t_k, i}^k)=\nabla f_k(w_{t_k, i}^k, \mathcal{D}_k)$. The server commences the aggregation process once \textit{K} devices have reported. We first outline the key assumptions necessary to present our theoretical analysis on the convergence of {\em SEAFL}, listed in the following.\\

\textbf{Assumption 1.} \textit{(Lipschitz gradient) The objective function of each device $f_k$ is L-smooth. Thus $f_k$ has Lipschitz continuous gradients with constant $L > 0$, i.e., 
$\|\nabla f_k(w) - \nabla f_k(w')\| \leq L \|w - w'\|$.}\\

\textbf{Assumption 2.} \textit{(Unbiased local gradient) For each device the stochastic gradient $ \nabla f_k(w;\xi)$ is unbiased, i.e.,
$\mathbb{E}[f_k(w;\xi] = \nabla f_k(w)$.} \\

\textbf{Assumption 3.} \textit{(Uniformly bounded local gradient) The expected squared norm of stochastic gradients is uniformly bounded, i.e.,
$\mathbb{E} \|\nabla f_k(w;\xi\|^2 \leq G^2$ for all $k = 1,\cdots,K$ and $t=1,\cdots,T-1.$}\\

\textbf{Assumption 4.} \textit{(Bounded local gradients) Let $\xi$ be a sample drawn uniformly at random from the local data of the k-th device. The variance of the stochastic gradients for each device is constrained as follows: $\mathbb{E}_\xi\|f_k(w;\xi)-f_k(w)\|^2 \leq \sigma_k^2$ for $k=1,\cdots,K$. We then define $\sigma_l^2:= \sum_{k=1}^{K} \frac{|\mathcal{D}_k|}{|D|} \sigma_k^2.$}\\

\textbf{Assumption 5.} \textit{(Bounded gradient dissimilarity) For any device k and parameter $w$, we denote $\delta_k$ as the upper bound for $\| f_k(w) - f(w)\|^2$, \smash{i.e., $\| f_k(w) - f(w)\|^2 \leq \delta_k^2$.} We then define $\delta_g^2:= \sum_{k=1}^{K} \frac{|\mathcal{D}_k|}{|D|} \delta_k^2.$}\\

{\em SEAFL} can be characterized as an asynchronous aggregation problem that incorporates buffered updates, a concept previously addressed in {\em FedBuff} \cite{nguyen2022federated}. Additionally, {\em SEAFL}'s partial training strategy also guarantees a staleness limit to device updates as mentioned in Section~\ref{sec:design}. Furthermore, we mathematically define a staleness factor that leverages the devices's staleness to modify the weights assigned to each gradient. By incorporating the importance factor, we establish Lemma 1 regarding the weights assigned to each gradient. \\

\textbf{Lemma 1.} \textit{Given the hyperparameters associated with the staleness factor and importance factor, $\alpha$ and $\mu$, the aggregation weight  $p_t^k$ for each gradient can be bounded within the interval $p_t^k\in[\frac{\alpha}{2}d_k, (\alpha+\mu)d_k]$ where \smash{$d_k=\frac{|\mathcal{D}_k|}{|D|}$.}}\\

We can simplify Lemma 1 by ignoring the denominator term since it does not influence the convergence proof. Consequently, we can derive the convergence rate of {\em SEAFL} as follows:\\

\textbf{Theorem 1} (Convergence rate) \textit{Based on Assumptions 1 to 4 and Lemma 1, the convergence rate of SEAFL can be formulated as follows:}
\begin{equation}
\begin{aligned}
\frac{1}{T}\sum_{t=0}^{T-1} \mathbb{E}\|\nabla f(w_t)\|^2 &\leq \frac{2 (f(w_0)-f(w^*))}{\Omega(E)TK} \\
&\hspace{-5em}+ 6K (\alpha + \mu)^2 \lambda (d)L^2 Q\phi (E) (K^2\beta^2+1)\sigma^2 \\
&\hspace{-5em}+ \frac{\phi(E)L}{K\Omega(E)} (\alpha+\mu)\sigma_l^2
\end{aligned}
\end{equation}
\textit{where $\Omega (E)=\sum_{i=1}^{E} n_l^i$, $\lambda(d)=\sum_{j=1}^{K} d_j^2$, $\phi(E)=\sum_{i=1}^{E} (\eta_l^i)^2$, and $\sigma^2= (\alpha+\mu)\sigma_l^2 + (\alpha+\mu)\sigma_g^2 + G^2.$}\\

\textit{To achieve the upper bound on convergence, the relationship between K and $n_l$ must satisfy the following condition:}
\begin{equation}
\frac{4(\alpha+\mu)}{\alpha^2 \lambda(d)}K\eta_l^i\leq \frac{1}{L}
\end{equation}

\textit{Proof}. Following the conventional method for proving convergence in federated learning algorithms, such as in \cite{nguyen2022federated}, which handles non-convex objective function, our proof begins by applying the smoothness Assumption 1. This allows us to establish an upper bound for $f(w_{t+1})$ as follows:
\begin{equation}
\begin{aligned}
f(w_{t+1}) \leq f(w_t) - \sum_{k\in K}p_t^k(\nabla f(w_t),\Delta_{t_k})\\
&\hspace{-8em}+ \frac{L}{2}\|\sum_{k\in K}p_t^k\Delta_{t_k}\|^2
\end{aligned}
\end{equation}
where $\Delta_{t_k} = \sum_{i=1}^{E}n_l^i\nabla f_k \left(w_{t_k, i}^k \right).$

Then, as presented in Eq. (7), {\em SEAFL} evaluates the staleness factor for each gradient. It incorporates a new aggregation mechanism that generalizes the scenario considered in {\em FedBuff} \cite{nguyen2022federated}, where equal weights are assigned during aggregation. Specifically, we outline our proof in three parts.

\setlength{\abovecaptionskip}{.5pt}
\begin{figure*}
\centering
\captionsetup[subfigure]{skip=-3pt}
{\includegraphics[height=5.2cm, width=0.32\textwidth]{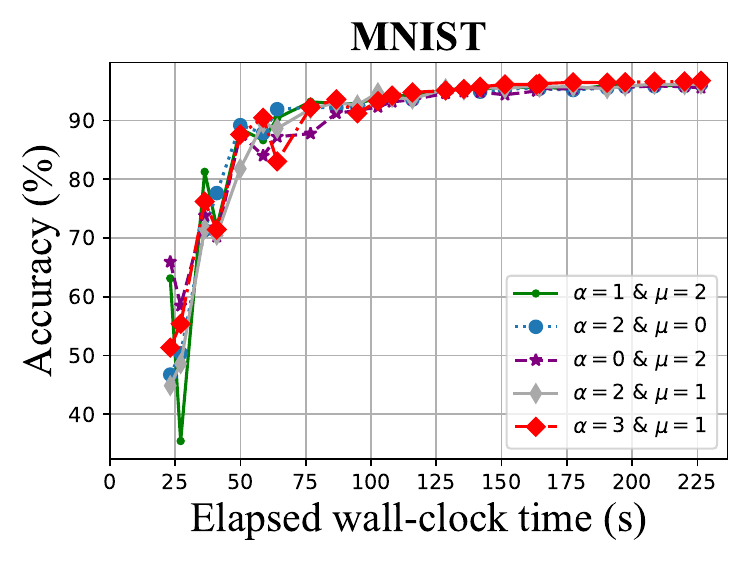}}
{\includegraphics[height=5.2cm, width=0.32\textwidth]{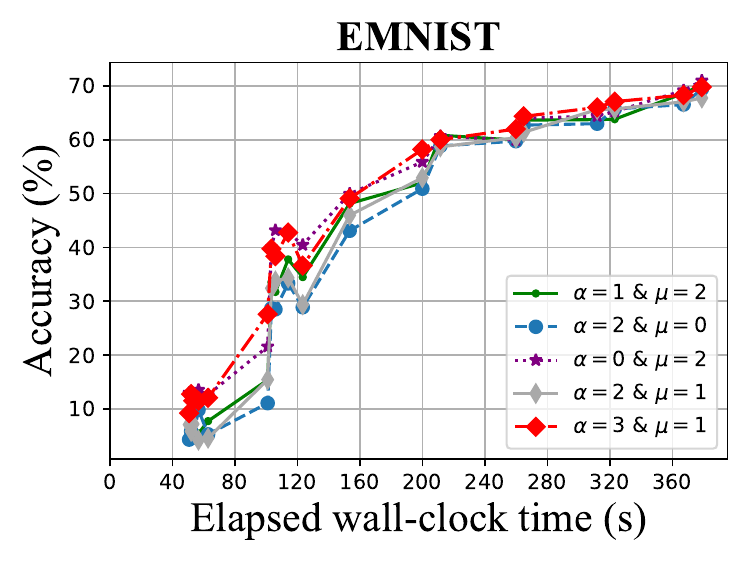}}
{\includegraphics[height=5.2cm, width=0.32\textwidth]{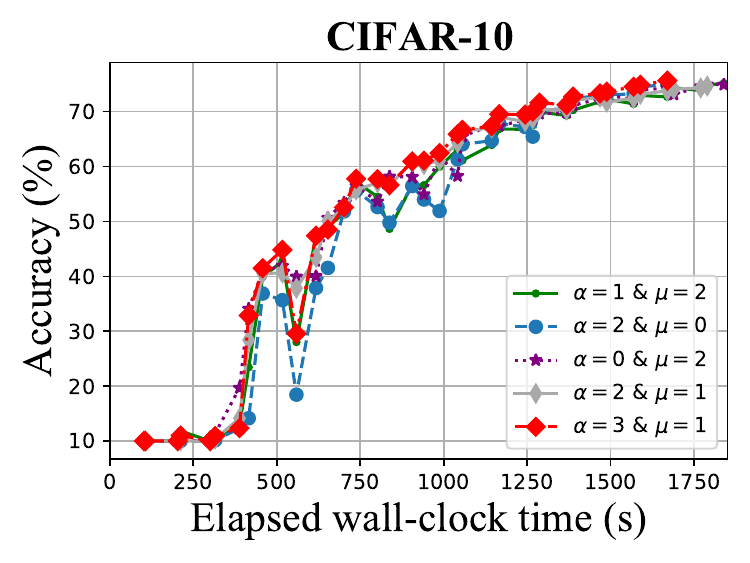}}
\caption{Elapsed wall-clock time required to reach target accuracy for different combinations of $\alpha$ and $\mu$.}
\label{figure4}
\end{figure*}

Initially, we derive the upper bound for three crucial components. According to Assumptions 3, 4, 5, and Lemma 1, we establish a bound on the expectation of the stochastic gradient $\mathbb{E}\|\nabla f_k (w_{t_k, i}^k, \mathcal{D}_k)\|$ of device \textit{k} by $\sigma^2= (\alpha+\mu)\sigma_l^2 + (\alpha+\mu)\sigma_g^2 + G^2.$  Next, we prove the upper bound for staleness-aware gradient divergence by utilizing Assumption 1 and including a zero term in the decomposition $\mathbb{E}\|\sum_{k=1}^{K}p_t^k(\nabla f_k (w_t)- f_k(w_{t_k}^k))\|^2$ is $6K\sum_{k=1}^{K}(p_t^k)^2 \sum_{k=1}^{K}L^2 Q\phi (E) (K^2\beta^2+1)\sigma^2 $. Finally, we establish a bound on $\mathbb{E}\|\sum_{k\in K}p_t^k\Delta_{t_k}\|^2$ by employing Lemma 1 and Assumption 5.

Then, incorporating these derived components into Eq. (11), we manipulate the equation to derive the specific upper bound for $\mathbb{E}[f(w_t)]$. To eliminate the term containing $\mathbb{E}\|\nabla f_k(w_{t_k}^k)\|^2$, we aim to make the upper bound of its coefficient to 0, i.e., $-\frac{K}{2}\left(\sum_{k=1}^{K}(p_t^k)^2\right)+\frac{LK^2E(n_l^i)^2}{2}p_t^k \leq 0$. Therefore, based on Lemma 1, we obtain Eq. (10). Finally, with the simplified right-hand side of Eq. (11), We compute the summation from 1 to \textit{T} and reorganize the equation to obtain Eq. (9).\\

We present Corollary 1, which is based on Theorem 1:\\

\textbf{Corollary 1.} \textit{In accordance with the convergence rate established in Theorem 1, when $n_l$ is a constant value and satisfies the conditions in Eq. (10), i.e., $n_l=\frac{1}{\sqrt{{TKE}}}$, then, we derive for a sufficiently large T:}

\begin{equation}
\begin{aligned}
\frac{1}{T}\sum_{t=0}^{T-1} \mathbb{E}\|\nabla f(w_t)\|^2 &\leq \mathcal{O} \left(\frac{(f(w_0)-f(w^*))}{\sqrt{TKE}}\right)\\
&+ \mathcal{O}  \left(\frac{EK^2\beta^2\sigma^2}{T}\right) + \mathcal{O} \left(\frac{E\sigma^2}{T} \right) \\
&+\mathcal{O} \left(\frac{\sigma_l^2}{K\sqrt{TKE}} \right)
\end{aligned}
\end{equation}
where $\sigma^2= (\alpha+\mu)\sigma_l^2 + (\alpha+\mu)\sigma_g^2 + G^2.$\\

The proof of this corollary is excluded due to space limitations. Our theoretical analysis allows us to highlight several critical observations regarding the key factors influencing convergence.

\textit{The staleness limit $\beta$.} The effect of the staleness limit on convergence diminishes at a rate of $1/T$, as indicated by the second term in Eq. (12). A large staleness limit is not preferable due to its contribution to an increase in the second term. Nevertheless, we can modify the buffer size to mitigate its influence on the convergence rate.

\textit{The buffer size \textit{K}.} The first term of Eq. (12) indicates that as the buffer size \textit{K} increases, there is a rapid decrease in the loss towards the optimal value. However, the impact of the variance $\sigma^2$ can be increased, thus enhancing the gradient drift during training. Furthermore, a large staleness limit $\beta$ results in server aggregation incorporating outdated updates while waiting for more devices, which negatively affect convergence. As a result, we anticipate $K\in (1, M] $, where \textit{M} should not be excessively large.

In contrast to related approaches such as {\em FedBuff} \cite{nguyen2022federated}, {\em SEAFL} represents a more generalized framework of semi-asynchronous aggregation techniques with buffered updates. {\em SEAFL}'s convergence rate naturally degenerates into Theorem 1 in \cite{nguyen2022federated} by setting consistent weights $p_t^k=\frac{1}{K}$ for gradients in the server aggregation process.

\section{Performance Evaluation}
\label{sec:results}
In this section, we present the experimental comparison of {\em SEAFL} with three state-of-the-art approaches across three commonly used datasets to validate its effectiveness.
\subsection{Experimental Setup}
\textbf{Datasets and Models.} 
The experiments are conducted over different image classification tasks using three popular benchmark datasets i.e., EMNIST \cite{cohen2017emnist}, CIFAR-10 \cite{krizhevsky2009learning}, and CINIC-10 \cite{darlow2018cinic}. The data distributed across each device exhibits a non-IID pattern, generated from a Dirichlet distribution with a concentration parameter of 5 for all datasets. In our experiments, we consider LeNet-5 \cite{lecun1989backpropagation} model for EMNIST, ResNet-18 \cite{he2016deep} for CIFAR-10, and VGG-16 \cite{simonyan2014very} for CINIC-10. 


\textbf{Baselines Methods.} 
To demonstrate the performance of {\em SEAFL}, we compare it against three state-of-the-art (SOTA) FL approaches: (i) {\em FedAvg} \cite{mcmahan2017communication}, which is a synchronous federated learning approach; (ii) {\em FedBuff} \cite{nguyen2022federated}, a semi-asynchronous method for federated learning; and (iii) {\em FedAsync} \cite{xie2019asynchronous}, the standard asynchronous federated learning approach.

\setlength{\abovecaptionskip}{.5pt}
\begin{figure*}
\centering
{\includegraphics[height=5.2cm, width=0.32\textwidth]{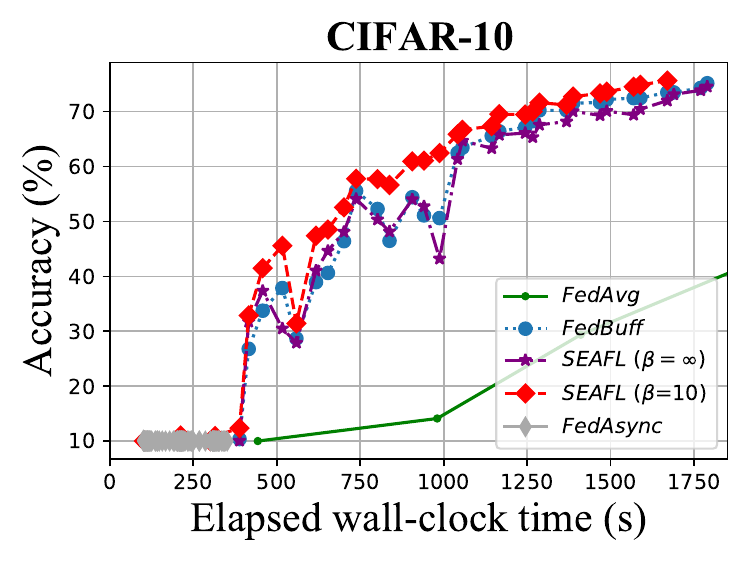}}
{\includegraphics[height=5.2cm, width=0.32\textwidth]{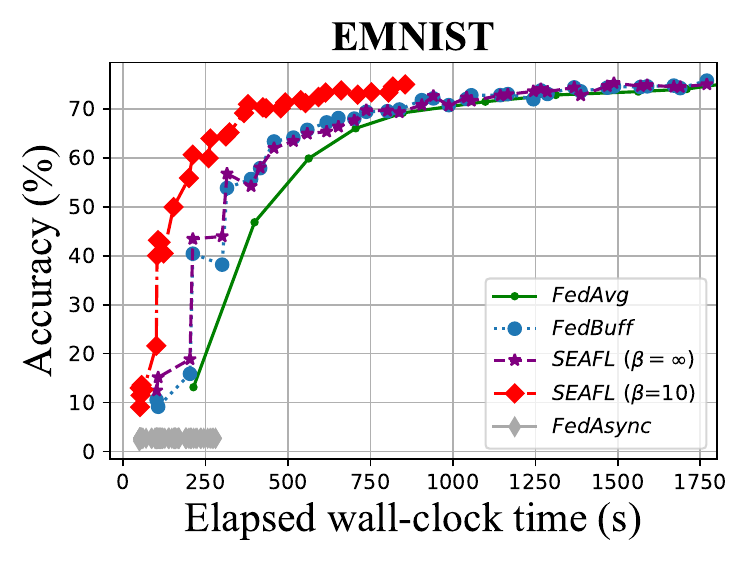}}
{\includegraphics[height=5.2cm, width=0.32\textwidth]{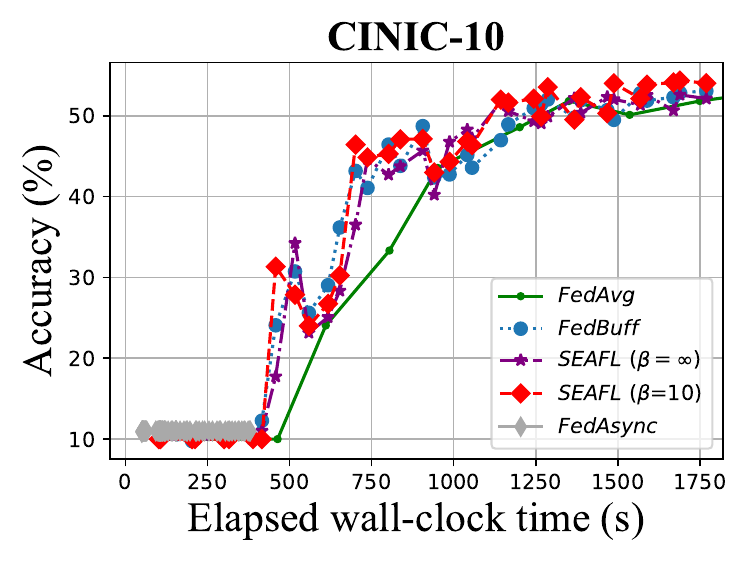}}
\caption{Elapsed wall-clock time required to reach target accuracy for {\em SEAFL} (without partial training), {\em FedBuff}, {\em FedAsync}, and {\em FedAvg}. {\em SEAFL} converges faster to reach target accuracy and consistently outperforms other baselines.}
\label{figure5}
\end{figure*}

\textbf{Implementation.} 
We have implemented {\em SEAFL} using the open-source research framework PLATO \cite{li2023plato}. This framework supports the emulation of asynchronous FL training and provides the ability to measure the elapsed wall-clock time during an FL training session. We assume 100 devices are available for all experiments and up to 20\% of them are sampled randomly in each communication round for the synchronous mode. In all experiments, we set $\vartheta=0.8$, and epochs $E=5$. All experiments are performed on a server equipped with an NVIDIA GeForce RTX 3080 Ti GPU. We used Pareto distribution to simulate heavy-tailed client speed.

\textbf{Evaluation Metrics.} 
Existing works typically evaluate performance based solely on metrics such as the number of gradients, updates, or communication rounds required to reach a target accuracy.  However, these metrics may not accurately reflect real-world training time due to the nature of asynchronous FL, where the communication round index can advance whenever a single device reports to the server. As a result, these metrics often fail to reflect the actual wall-clock time needed to achieve target accuracy. Hence, we consider the elapsed wall-clock time required to reach a target accuracy on the test set, rather than the number of rounds.

\subsection{Results and Analysis}

\textbf{Effect of hyperparameters.}
Initially, we run a large number of experiments to figure out the optimal combination of hyperparameters $\alpha$ and $\mu$, crucial for our adaptive weight aggregation mechanism. We explore values ranging from 0 to 10 for both $\alpha$ and $\mu$. Fig.~\ref{figure4} illustrates the comparison of various representative pairs of values for $\alpha$ and $\mu$. In general, the combination of $\alpha=3$  and $\mu=1$ provided a modest performance improvement compared to other value pairs.

\textbf{\textit{SEAFL} vs. baselines.}
We compare the performance of {\em SEAFL} (without partial training) with baseline methods. In Fig.~\ref{figure5}, we present the wall-clock training time required to achieve a target accuracy for each dataset. It is evident that {\em FedAsync} completely failed to converge in all cases. This is attributed to its aggressive strategy of immediately aggregating the fastest device update upon arrival, as well as the design of its own aggregation algorithm. {\em SEAFL} consistently outperformed the synchronous FL baseline, {\em FedAvg}, in terms of the wall-clock training times needed to reach target accuracy for all datasets. 
\setlength{\abovecaptionskip}{.5pt}
\begin{figure}
\centering
\captionsetup[subfigure]{skip=-3pt}
\subfloat[\textit{SEAFL$^2$} with staleness limit of 3 vs. baselines]
{\includegraphics[height=5.2cm, width=0.32\textwidth]{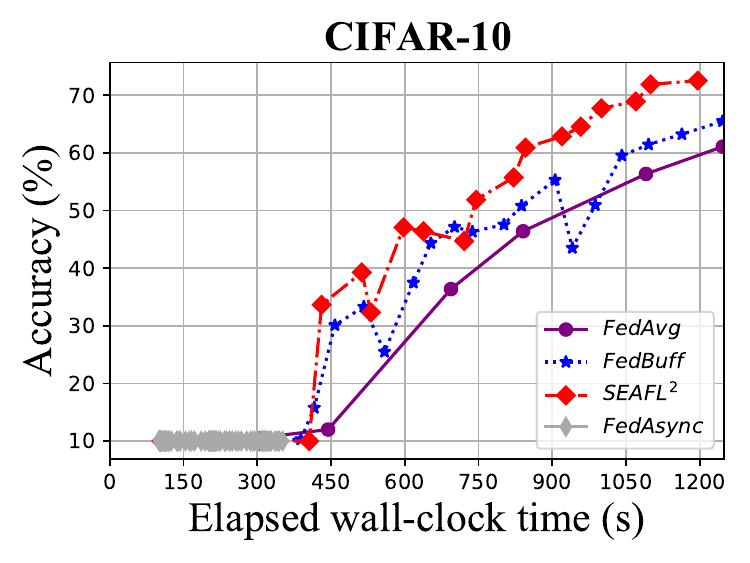}\label{fig:1}}
\hspace{0.05\textwidth} 
\subfloat[\textit{SEAFL$^2$} with staleness limit of 12 vs. baselines]
{\includegraphics[height=5.2cm, width=0.32\textwidth]{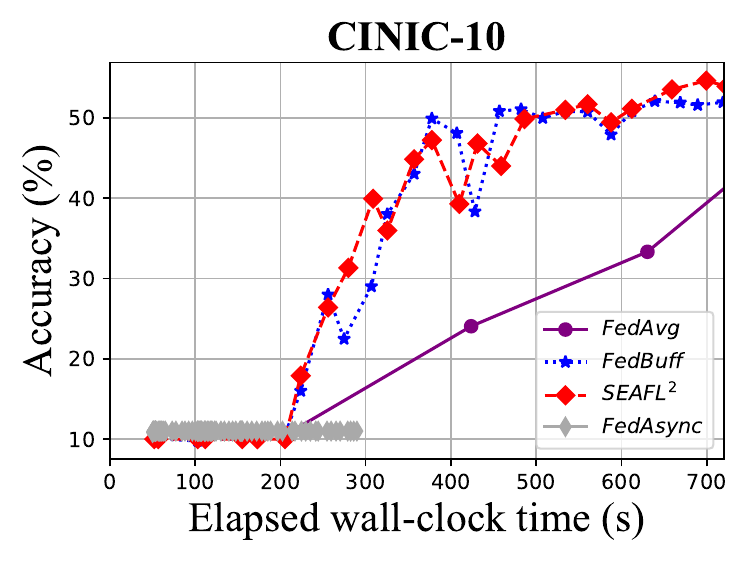}\label{fig:2}}

\caption{Performance comparison of \textit{SEAFL$^2$} and baseline methods.}
\label{figure6}
\end{figure}

As {\em FedBuff} does not impose any restriction on staleness, it effectively operates with an $\infty$ staleness limit. Therefore, we conducted experiments with all three scenarios: {\em SEAFL} with a staleness limit of 10, {\em SEAFL} with an $\infty$ staleness limit, and {\em FedBuff}. From Fig.~\ref{figure5}, we can see that, {\em SEAFL} performed very similarly while having an $\infty$ staleness limit. This is attributed to the fact that the majority of the aggregated device updates are not too stale. However, both approaches suffer from accuracy degradation for a few rounds when stale devices eventually arrive. In contrast, {\em SEAFL} exhibits superior performance across all datasets with a staleness limit of 10, especially with EMNIST. These experiments support the intuition that imposing a reasonable staleness limit provides benefits in terms of wall-clock time, even while accommodating slower devices.

\textbf{{\em SEAFL$^2$} vs. baselines.}
In this series of experiments, we activate the partial training mechanism in {\em SEAFL} and measure the resulting training durations. Our findings are depicted in Fig.~\ref{figure6}. With a low staleness limit of 3, {\em SEAFL$^2$} enabled the server to promptly notify devices upon reaching the staleness limit while working with the CIFAR-10 dataset. In this scenario, {\em SEAFL$^2$} has clearly shown its advantages: it achieved 50\% accuracy in only 745 seconds, and reached 70\% accuracy in 1105 seconds. In contrast, its closest rival, {\em FedBuff}, required 905 seconds to achieve 50\% and 1341 seconds to achieve 70\%. This indicates that {\em SEAFL$^2$} achieved a performance advantage of up to 22\% compared to {\em FedBuff}.

We experimented with a higher staleness limit of 12 using the CINIC-10 dataset to observe the performance of {\em SEAFL$^2$}. Fig.~\ref{fig:2} depicts that it initially progresses similarly to {\em FedBuff}, and only shows a slight advantage as convergence approaches completion. With the CINIC-10 dataset, where each device utilized only 3\% of the total samples for training compared to 10\% with CIFAR-10, this observation suggests that a staleness limit of $\infty$ may not be detrimental at all if local training finishes quickly and there is a high turnover rate to new devices. Consequently, in such scenarios, the performance benefit of {\em SEAFL$^2$} over {\em FedBuff} may decrease, as the impacts of partial training on stale devices become less impactful. However, in contrast to \textit{FedAsync} and \textit{FedAvg}, it is evident that {\em SEAFL$^2$} exhibits a significant performance advantage with both datasets.

Finally, it is important to highlight that, unlike the fully asynchronous operation with an $\infty$ staleness limit employed in {\em FedBuff}, a finite staleness limit offers a well-established and appealing theoretical property: guaranteed convergence during training \cite{ho2013more}. Despite the consistent convergence of {\em FedBuff} in our experiments, having a theoretical guarantee provides further assurance.

\section{Conclusion}
\label{sec:end}
In this work, we designed a novel staleness-aware semi-asynchronous FL framework, i.e., {\em SEAFL}, to address the straggler and excessive computation overhead issues of synchronous and asynchronous FL. {\em SEAFL} adaptively assigns weights to local updates during aggregation, considering both the staleness of received model updates and their similarity to the current global model. Additionally, we proposed {\em SEAFL$^2$} to further enhance the training efficiency, facilitating partial training on slower devices. {\em SEAFL$^2$} allows straggler devices to contribute to global aggregation and reduces the overall waiting time. Experimental results demonstrated significant advantages of {\em SEAFL} over state-of-the-art synchronous and asynchronous counterparts in terms of the convergence time required to achieve target accuracy. Moreover, we provided the theoretical convergence analysis of our proposed approach. In the future, we plan to extend {\em SEAFL} by incorporating adaptive partial training to dynamically adjust local model sizes using sub-model extraction techniques tailored to devices' real-time resource capabilities.

\bibliographystyle{IEEEtran}
\bibliography{sample-base}

\begin{thebibliography}{10}
\providecommand{\url}[1]{#1}
\csname url@samestyle\endcsname
\providecommand{\newblock}{\relax}
\providecommand{\bibinfo}[2]{#2}
\providecommand{\BIBentrySTDinterwordspacing}{\spaceskip=0pt\relax}
\providecommand{\BIBentryALTinterwordstretchfactor}{4}
\providecommand{\BIBentryALTinterwordspacing}{\spaceskip=\fontdimen2\font plus
\BIBentryALTinterwordstretchfactor\fontdimen3\font minus \fontdimen4\font\relax}
\providecommand{\BIBforeignlanguage}[2]{{%
\expandafter\ifx\csname l@#1\endcsname\relax
\typeout{** WARNING: IEEEtran.bst: No hyphenation pattern has been}%
\typeout{** loaded for the language `#1'. Using the pattern for}%
\typeout{** the default language instead.}%
\else
\language=\csname l@#1\endcsname
\fi
#2}}
\providecommand{\BIBdecl}{\relax}
\BIBdecl

\bibitem{gdpr}
\BIBentryALTinterwordspacing
{ EU. 2018. European Union's General Data Protection Regulation (GDPR)}. {European Union}. Accessed 2024-04. [Online]. Available: \url{https://eugdpr.org/}
\BIBentrySTDinterwordspacing

\bibitem{mcmahan2017communication}
B.~McMahan, E.~Moore, D.~Ramage, S.~Hampson, and B.~A. y~Arcas, ``Communication-efficient learning of deep networks from decentralized data,'' in \emph{Artificial intelligence and statistics}.\hskip 1em plus 0.5em minus 0.4em\relax PMLR, 2017, pp. 1273--1282.

\bibitem{liu2021federated}
M.~Liu, S.~Ho, M.~Wang, L.~Gao, Y.~Jin, and H.~Zhang, ``Federated learning meets natural language processing: A survey,'' \emph{arXiv preprint arXiv:2107.12603}, 2021.

\bibitem{liu2020fedvision}
Y.~Liu, A.~Huang, Y.~Luo, H.~Huang, Y.~Liu, Y.~Chen, L.~Feng, T.~Chen, H.~Yu, and Q.~Yang, ``Fedvision: An online visual object detection platform powered by federated learning,'' in \emph{Proceedings of the AAAI Conference on Artificial Intelligence}, vol.~34, no.~08, 2020, pp. 13\,172--13\,179.

\bibitem{nguyen2022federated}
J.~Nguyen, K.~Malik, H.~Zhan, A.~Yousefpour, M.~Rabbat, M.~Malek, and D.~Huba, ``Federated learning with buffered asynchronous aggregation,'' in \emph{International Conference on Artificial Intelligence and Statistics}.\hskip 1em plus 0.5em minus 0.4em\relax PMLR, 2022, pp. 3581--3607.

\bibitem{ouyang2021clusterfl}
X.~Ouyang, Z.~Xie, J.~Zhou, J.~Huang, and G.~Xing, ``Clusterfl: a similarity-aware federated learning system for human activity recognition,'' in \emph{Proceedings of the 19th Annual International Conference on Mobile Systems, Applications, and Services}, 2021, pp. 54--66.

\bibitem{jiang2022pisces}
Z.~Jiang, W.~Wang, B.~Li, and B.~Li, ``Pisces: Efficient federated learning via guided asynchronous training,'' in \emph{Proceedings of the 13th Symposium on Cloud Computing}, 2022, pp. 370--385.

\bibitem{lai2021oort}
F.~Lai, X.~Zhu, H.~V. Madhyastha, and M.~Chowdhury, ``Oort: Efficient federated learning via guided participant selection,'' in \emph{Proceedings of the Symposium on Operating Systems Design and Implementation}, 2021, pp. 19--35.

\bibitem{yang2021characterizing}
C.~Yang, Q.~Wang, M.~Xu, Z.~Chen, K.~Bian, Y.~Liu, and X.~Liu, ``Characterizing impacts of heterogeneity in federated learning upon large-scale smartphone data,'' in \emph{Proceedings of the Web Conference 2021}, 2021, pp. 935--946.

\bibitem{wu2020safa}
W.~Wu, L.~He, W.~Lin, R.~Mao, C.~Maple, and S.~Jarvis, ``Safa: A semi-asynchronous protocol for fast federated learning with low overhead,'' \emph{IEEE Transactions on Computers}, vol.~70, no.~5, pp. 655--668, 2020.

\bibitem{xie2019asynchronous}
C.~Xie, S.~Koyejo, and I.~Gupta, ``Asynchronous federated optimization,'' \emph{arXiv preprint arXiv:1903.03934}, 2019.

\bibitem{xu2023asynchronous}
C.~Xu, Y.~Qu, Y.~Xiang, and L.~Gao, ``Asynchronous federated learning on heterogeneous devices: A survey,'' \emph{Computer Science Review}, vol.~50, p. 100595, 2023.

\bibitem{zhou2022efficient}
C.~Zhou, J.~Liu, J.~Jia, J.~Zhou, Y.~Zhou, H.~Dai, and D.~Dou, ``Efficient device scheduling with multi-job federated learning,'' in \emph{Proceedings of the AAAI Conference on Artificial Intelligence}, vol.~36, no.~9, 2022, pp. 9971--9979.

\bibitem{zhou2024asynchronous}
Y.~Zhou, X.~Pang, Z.~Wang, J.~Hu, P.~Sun, and K.~Ren, ``Towards efficient asynchronous federated learning in heterogeneous edge environments,'' in \emph{Proceedings of the IEEE Conference on Computer Communications (INFOCOM)}, 2024.

\bibitem{liu2023fedasmu}
J.~Liu, J.~Jia, T.~Che, C.~Huo, J.~Ren, Y.~Zhou, H.~Dai, and D.~Dou, ``Fedasmu: Efficient asynchronous federated learning with dynamic staleness-aware model update,'' \emph{arXiv preprint arXiv:2312.05770}, 2023.

\bibitem{wang2022asyncfeded}
Q.~Wang, Q.~Yang, S.~He, Z.~Shi, and J.~Chen, ``Asyncfeded: Asynchronous federated learning with euclidean distance based adaptive weight aggregation,'' \emph{arXiv preprint arXiv:2205.13797}, 2022.

\bibitem{chen2020asynchronous}
Y.~Chen, Y.~Ning, M.~Slawski, and H.~Rangwala, ``Asynchronous online federated learning for edge devices with non-iid data,'' in \emph{Proceedings of IEEE International Conference on Big Data (Big Data)}.\hskip 1em plus 0.5em minus 0.4em\relax IEEE, 2020, pp. 15--24.

\bibitem{li20201federated}
T.~Li, A.~K. Sahu, M.~Zaheer, M.~Sanjabi, A.~Talwalkar, and V.~Smith, ``Federated optimization in heterogeneous networks,'' \emph{Proceedings of Machine Learning and Systems}, vol.~2, pp. 429--450, 2020.

\bibitem{li2021ditto}
T.~Li, S.~Hu, A.~Beirami, and V.~Smith, ``Ditto: Fair and robust federated learning through personalization,'' in \emph{Proceedings of International Conference on Machine Learning}.\hskip 1em plus 0.5em minus 0.4em\relax PMLR, 2021, pp. 6357--6368.

\bibitem{zhang2023fedala}
J.~Zhang, Y.~Hua, H.~Wang, T.~Song, Z.~Xue, R.~Ma, and H.~Guan, ``Fedala: Adaptive local aggregation for personalized federated learning,'' in \emph{Proceedings of the AAAI Conference on Artificial Intelligence}, vol.~37, no.~9, 2023, pp. 11\,237--11\,244.

\bibitem{ghosh2020efficient}
A.~Ghosh, J.~Chung, D.~Yin, and K.~Ramchandran, ``An efficient framework for clustered federated learning,'' \emph{Advances in Neural Information Processing Systems}, vol.~33, pp. 19\,586--19\,597, 2020.

\bibitem{islam2024fedclust}
M.~S. Islam, S.~Javaherian, F.~Xu, X.~Yuan, L.~Chen, and N.-F. Tzeng, ``Fedclust: Tackling data heterogeneity in federated learning through weight-driven client clustering,'' in \emph{Proceedings of the 53rd International Conference on Parallel Processing}, 2024, pp. 474--483.

\bibitem{javaherian2024fedfair}
S.~Javaherian, S.~Panta, S.~Williams, M.~S. Islam, and L.~Chen, ``Fedfair\^{}3: Unlocking threefold fairness in federated learning,'' \emph{in Proceedings of IEEE International Conference on Communications (ICC)}, pp. 1--7, 2024.

\bibitem{li2022pyramidfl}
C.~Li, X.~Zeng, M.~Zhang, and Z.~Cao, ``Pyramidfl: A fine-grained client selection framework for efficient federated learning,'' in \emph{Proceedings of the 28th Annual International Conference on Mobile Computing And Networking}, 2022, pp. 158--171.

\bibitem{zhang2022fedduap}
H.~Zhang, J.~Liu, J.~Jia, Y.~Zhou, H.~Dai, and D.~Dou, ``Fedduap: Federated learning with dynamic update and adaptive pruning using shared data on the server,'' \emph{arXiv preprint arXiv:2204.11536}, 2022.

\bibitem{horvath2021fjord}
S.~Horvath, S.~Laskaridis, M.~Almeida, I.~Leontiadis, S.~Venieris, and N.~Lane, ``Fjord: Fair and accurate federated learning under heterogeneous targets with ordered dropout,'' \emph{Advances in Neural Information Processing Systems}, vol.~34, pp. 12\,876--12\,889, 2021.

\bibitem{li2022fedhisyn}
G.~Li, Y.~Hu, M.~Zhang, J.~Liu, Q.~Yin, Y.~Peng, and D.~Dou, ``Fedhisyn: A hierarchical synchronous federated learning framework for resource and data heterogeneity,'' in \emph{Proceedings of the 51st International Conference on Parallel Processing}, 2022, pp. 1--11.

\bibitem{abad2020hierarchical}
M.~S.~H. Abad, E.~Ozfatura, D.~Gunduz, and O.~Ercetin, ``Hierarchical federated learning across heterogeneous cellular networks,'' in \emph{Proceedings of ICASSP 2020-2020 IEEE International Conference on Acoustics, Speech and Signal Processing (ICASSP)}.\hskip 1em plus 0.5em minus 0.4em\relax IEEE, 2020, pp. 8866--8870.

\bibitem{zhang2023timelyfl}
T.~Zhang, L.~Gao, S.~Lee, M.~Zhang, and S.~Avestimehr, ``Timelyfl: Heterogeneity-aware asynchronous federated learning with adaptive partial training,'' in \emph{Proceedings of the IEEE/CVF Conference on Computer Vision and Pattern Recognition}, 2023, pp. 5063--5072.

\bibitem{ma2021fedsa}
Q.~Ma, Y.~Xu, H.~Xu, Z.~Jiang, L.~Huang, and H.~Huang, ``Fedsa: A semi-asynchronous federated learning mechanism in heterogeneous edge computing,'' \emph{IEEE Journal on Selected Areas in Communications}, vol.~39, no.~12, pp. 3654--3672, 2021.

\bibitem{li2022federated}
Q.~Li, Y.~Diao, Q.~Chen, and B.~He, ``Federated learning on non-iid data silos: An experimental study,'' in \emph{Proceedings of IEEE 38th International Conference on Data Engineering (ICDE)}.\hskip 1em plus 0.5em minus 0.4em\relax IEEE, 2022, pp. 965--978.

\bibitem{li2023plato}
B.~Li, N.~Su, C.~Ying, and F.~Wang, ``Plato: An open-source research framework for production federated learning,'' in \emph{Proceedings of the ACM Turing Award Celebration Conference-China 2023}, 2023, pp. 1--2.

\bibitem{cohen2017emnist}
G.~Cohen, S.~Afshar, J.~Tapson, and A.~Van~Schaik, ``Emnist: Extending mnist to handwritten letters,'' in \emph{Proceedings of International Joint Conference on Neural Networks (IJCNN)}.\hskip 1em plus 0.5em minus 0.4em\relax IEEE, 2017, pp. 2921--2926.

\bibitem{krizhevsky2009learning}
A.~Krizhevsky, G.~Hinton \emph{et~al.}, ``Learning multiple layers of features from tiny images,'' 2009.

\bibitem{darlow2018cinic}
L.~N. Darlow, E.~J. Crowley, A.~Antoniou, and A.~J. Storkey, ``Cinic-10 is not imagenet or cifar-10,'' \emph{arXiv preprint arXiv:1810.03505}, 2018.

\bibitem{lecun1989backpropagation}
Y.~LeCun, B.~Boser, J.~S. Denker, D.~Henderson, R.~E. Howard, W.~Hubbard, and L.~D. Jackel, ``Backpropagation applied to handwritten zip code recognition,'' \emph{Neural Computation}, vol.~1, no.~4, pp. 541--551, 1989.

\bibitem{he2016deep}
K.~He, X.~Zhang, S.~Ren, and J.~Sun, ``Deep residual learning for image recognition,'' in \emph{Proceedings of the IEEE conference on computer vision and pattern recognition}, 2016, pp. 770--778.

\bibitem{simonyan2014very}
K.~Simonyan and A.~Zisserman, ``Very deep convolutional networks for large-scale image recognition,'' \emph{arXiv preprint arXiv:1409.1556}, 2014.

\bibitem{ho2013more}
Q.~Ho, J.~Cipar, H.~Cui, S.~Lee, J.~K. Kim, P.~B. Gibbons, G.~A. Gibson, G.~Ganger, and E.~P. Xing, ``More effective distributed ml via a stale synchronous parallel parameter server,'' \emph{Proceedings of Advances in Neural Information Processing Systems}, vol.~26, 2013.

\end{thebibliography}

\end{document}